\documentclass[a4paper,11pt]{article}
\usepackage{pos}

\DeclareMathOperator{\erf}{erf}

\newcommand{\s}{{\bf{s}}}
\newcommand{\Tc}{T_{\mbox{\tiny{c}}}}
\newcommand{\redchisq}{\chi^2_{\tiny\mbox{red}}}

\title{Monopole-like configurations in the O(3) spin model at the upper critical dimension}
 \ShortTitle{Topological objects in the O(3) spin model}

\author[]{Marco Panero}
\author*[]{Antonio Smecca}

\affiliation[]{Department of Physics, University of Turin,\\
  Via Pietro Giuria 1, 10125 Turin, Italy}

\affiliation[]{INFN, Turin,\\
Via Pietro Giuria 1, 10125 Turin, Italy}

\emailAdd{marco.panero@unito.it}
\emailAdd{antonio.smecca@unito.it}

\abstract{We present a high-precision Monte Carlo study of the $O(3)$ spin theory on the lattice in $D=4$ dimensions. This model exhibits interesting dynamical features, in particular in the broken-symmetry phase, where suitable boundary conditions can be used to enforce monopole-like topological excitations. We investigate the Euclidean time propagation and the features of these excitations close to the critical point, where our numerical results show an excellent quantitative agreement with analytic predictions derived from purely quantum-field-theoretical tools by G. Delfino. 
We conclude by commenting on the implications of our findings for a conjectured violation of Derrick's theorem at the quantum level and on the consequences in various areas of physics, ranging from condensed matter to astro-particle physics.}

\FullConference{%
 The 38th International Symposium on Lattice Field Theory, LATTICE2021
  26th-30th July, 2021
  Zoom/Gather@Massachusetts Institute of Technology
}

\begin{document}
\maketitle

\section{Introduction and Motivation}

In this contribution we present our latest work \cite{Papero}, regarding the study of the classical Heisenberg model at the upper critical dimension. This model experiences a phase transition at the critical temperature, allowing us to distinguish two phases in which the overall symmetry of the system is either spontaneously broken or restored \cite{Frolich}. In particular, the low-temperature phase encodes a non-trivial dynamics which allows us to study numerically non-local, finite-energy excitations, \textit{i.e.} topological defects, and to show that their properties can be successfully predicted using quantum-field-theoretical tools in a continuum formulation of the model as presented in \cite{Delfino}. For lower-dimensional systems, the theoretical expectations derived in that work have been qualitatively confirmed both analytically and numerically \cite{Berg, Delfino_2, XY}. Following these studies, our work gives an important contribution supporting the qualitative confirmation of these predictions considering that, for the first time, we numerically confirm these analytic predictions in four dimensions through Monte Carlo simulations in Euclidean spacetime lattice. The results obtained in this study gave us precious insight on the scaling properties of these topological objects, which, when compared with the results in the XY model \cite{XY}, allow us to check for possible violations of Derrick's theorem. Finally, the evidence collected from these theoretical studies on the existence of finite-energy excitations in vector spin models can find applications in other areas of physics, such as condensed matter physics and quantum simulation \cite{MagneticMonopoles, QuantumSimulation}.

\section{The O(3) model}
Lattice models of interacting vector spins with global symmetry $O(N)$, considering only nearest neighbours, can be described by the following Hamiltonian:

\begin{equation}
\mathcal{H} = - J \sum_{\langle x, y \rangle} \s(x) \cdot \s(y).
\end{equation}

Where $J$ is the ferromagnetic coupling and $\s(x)$ is a $N$-dimensional real vector of unit length, defined on a site $x$ of the Euclidean lattice in $D$ dimensions. One can identify particular cases for this model: the Ising model (for $N=1$), the XY model (for $N=2$) and the classical Heisenberg model (for $N=3$). Here we report our study on the Heisenberg model at the upper critical dimension, \textit{i.e.} $D=4$. As mentioned in the previous section, the classical Heisenberg model at the critical temperature goes through a phase transition that separates the ordered low-temperature phase from the disordered high-temperature one. At this phase transition, the critical exponents are equal to their mean field value, up to logarithmic corrections \cite{Kenna, Bauerschmidt}, which then allow us to study the system through Euclidean correlation functions.

\section{Computation Setup}

We work in a four-dimensional, isotropic, hyper-cubic lattice of spacing $a$. We denote the spatial extent of the system as $L$ and the temporal extent as $R$. We denote $T$ as the system temperature, and introduce the reduced temperature $t = (T-T_c)/T_c$, with $T_c$ being the critical temperature. We use the most recent estimate of the critical temperature, as determined in \cite{JPLv}.
In our Monte Carlo simulations, Markov chains of vector-field configurations are generated by a combination of local heath bath \cite{HeatBath_1, HeatBath_2, HeatBath_3} and overrelaxation updates \cite{Overrelaxation}.
In the high-temperature phase periodic boundary conditions are assumed. In this phase we consider the zero-spatial-momentum spin operators and extract the mass $m$ of the lightest physical state by fitting the two-point correlation function

\begin{equation}
  G(\tau,R) = \frac{a}{R}\sum_{x_0} {\bf{S}}(x_0) \cdot {\bf{S}}(x_0+\tau)
  \label{eq:two-point_corr}
\end{equation}
to the functional
\begin{equation}
  G(\tau,R) = A \left\{ \exp \left( -m \tau \right) + \exp\left[-m \left( R - \tau \right) \right] \right\}
  \label{eq:two-point_func}
\end{equation}
for sufficiently large $\tau$.

Switching to the low-temperature phase, we impose boundary conditions enforcing the existence of a ``monopole-like'' spin configuration. This is done by taking advantage of the topological nature that this type of excitations have in the continuum. Specifically, one can notice how the spatial boundary of the system at fixed Euclidean time and the manifold of the (classical) vacua share the same topology of the $S^2$ sphere, and use it to map these two manifolds in order to create the topological excitations. The simplest non-trivial mapping of this sort is the one that identifies the direction of $\s(x)$ with the direction of the spatial component of $x$ for all points at the spatial boundary with respect to the centre of the system. The same is done at the boundaries of the Euclidean time extent, which allow us to interpret the partition function $Z$ as the probability amplitude for the monopole to propagate from $x_0 = -R/2$ to $x_0 = R/2$.

\section{Results}

\subsection{High-Temperature Phase}
In the disordered high temperature phase our numerical results confirm the results obtained in \cite{JPLv}.

From our numerical simulations, we extract the masses of the excitations by fitting the two point correlation functions (eq. (\ref{eq:two-point_corr})) to the functional in eq. (\ref{eq:two-point_func}). We then perform a fit taking into account the results obtained with different reduced temperatures in order to obtain their values extrapolated to the thermodynamic limit. Neglecting logarithmic corrections, the masses extrapolated to the thermodynamic limit are expected to depend on $t$ as

\begin{equation}
  am = \frac{mt^{\nu}}{\Lambda_+}
  \label{TLimit}
\end{equation}
where $\Lambda_+$ is a constant with dimension of an energy, and the critical exponent is predicted to be $\nu = 1/2$, that is near the Gau{\ss}ian fixed point.

\begin{figure}[h!]
  \centering
  \includegraphics[height=0.3\textheight]{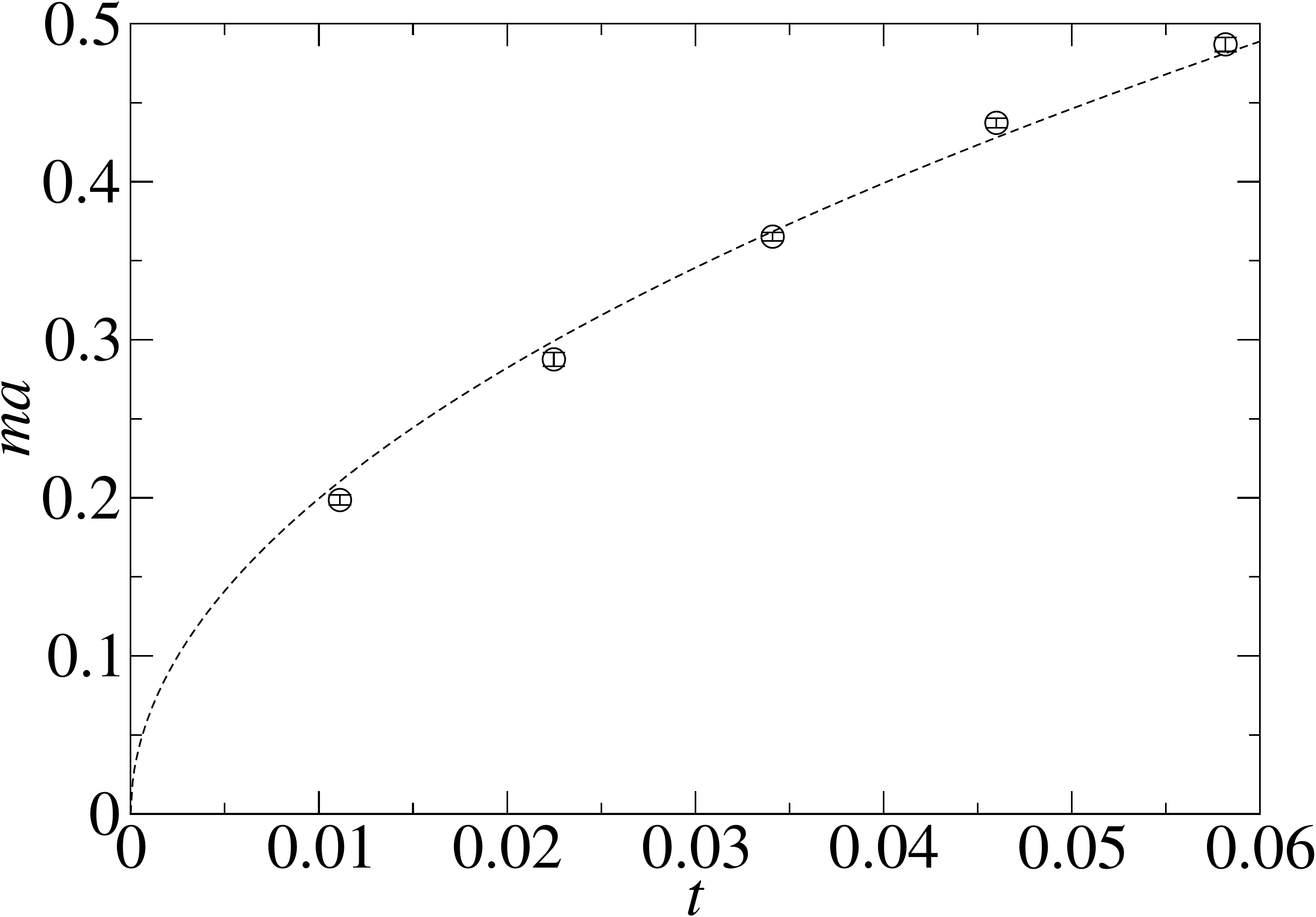}
  \caption{Mass values in units of the inverse lattice spacing at different reduced temperatures,
    extrapolated to the thermodynamic limit (circles), and their fit according to eq. (\ref{TLimit}) (dashed line)}
  \label{fig:TermodynamicLimit}
\end{figure}

Fitting our results to eq.~(\ref{TLimit}) we obtain the plot shown in figure \ref{fig:TermodynamicLimit}, which indicates that the mass of the lightest physical state that propagates in the theory vanishes as we approach the thermodynamic limit $T \rightarrow T_c$.

\subsection{Broken Symmetry Phase}

Our results in the low-temperature phase follow closely the analytical predictions in \cite{Delfino}, where spin models are studied through field theory arguments alone. Our results represent a first ever numerical confirmations in four Euclidean dimensions of the analytical predictions. In this phase we look at the spin profile and energy density in the presence of a topological defect enforced by the boundary conditions.

\begin{figure*}[]
\centerline{\includegraphics[height=0.26\textheight]{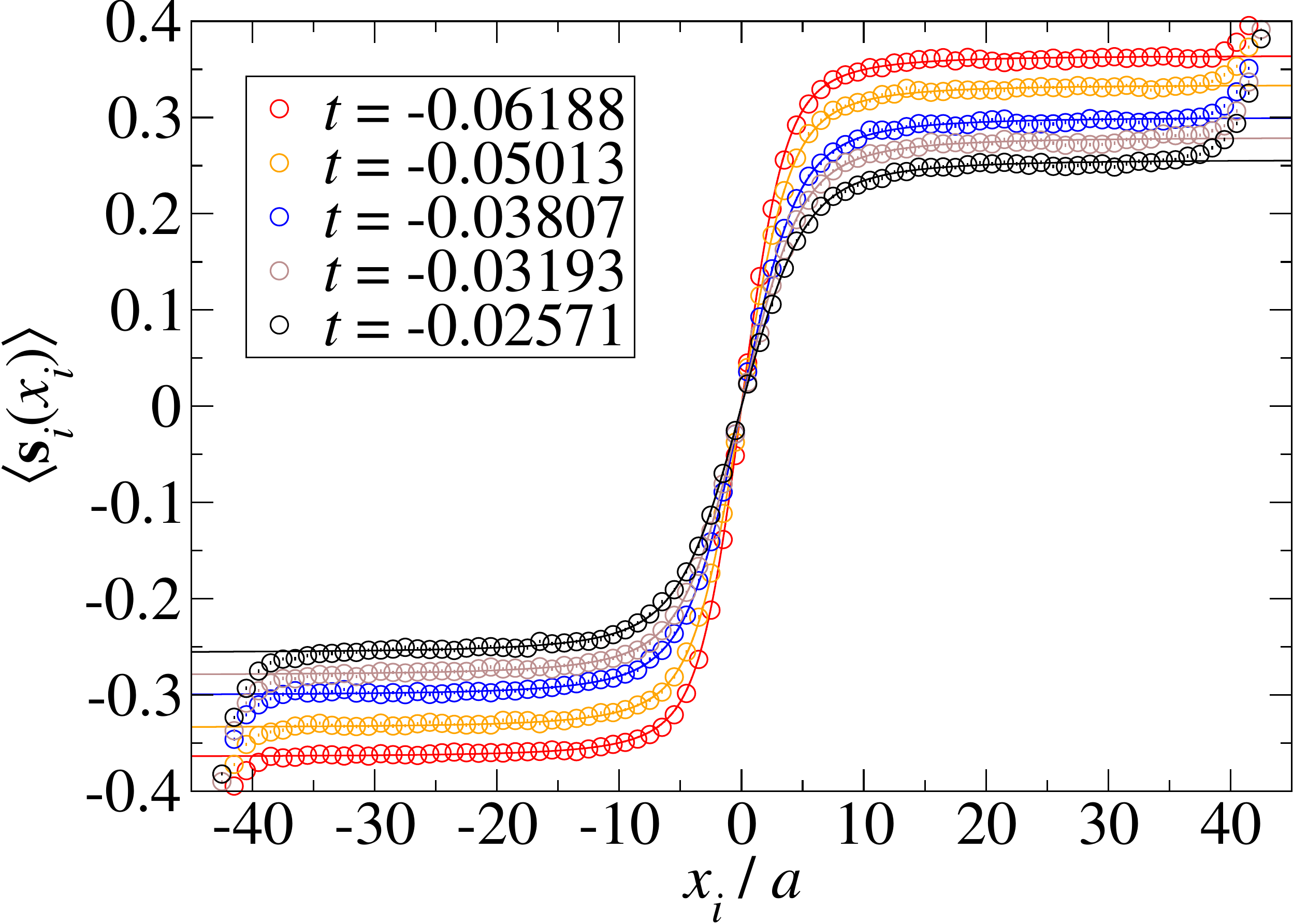} \hfill \includegraphics[height=0.26\textheight]{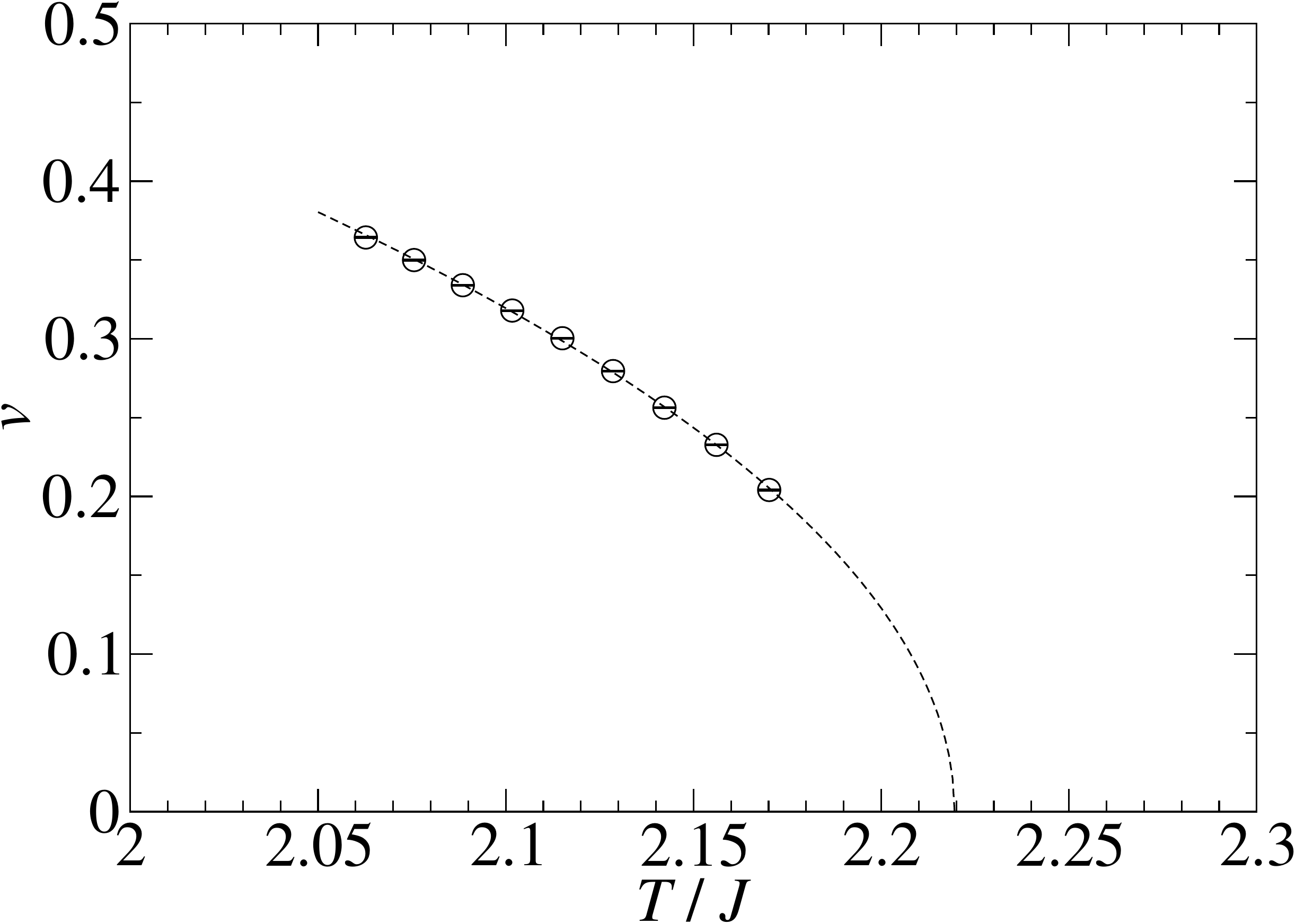}}
\caption{Left-hand-side panel: profile of the $i$-th component of the spin along the main axes through the center of the lattice, as a function of the $x_i$ coordinate (in units of the lattice spacing), in the presence of boundary conditions enforcing a topological excitation in the low-temperature phase. The plot shows the results obtained from our Monte~Carlo simulations on a lattice of spatial sizes $L=90a$ and Euclidean-time extent $R=20a$, denoted as circles of different colours, for different values of the reduced temperature $t$, and their comparison with the theoretical expectation according to eq.~(\ref{eq:SpinProfile}), which was derived in ref.~\cite{Delfino} from quantum-field-theory arguments. Right-hand-side panel: the average spin value $v$ far from the defect core, obtained from fits to eq.~(\ref{eq:SpinProfile}), plotted against the temperature in units of the coupling, from different sets of simulations on lattices with $L=90a$ and $R=20a$ (including those displayed in the plot on the right-hand-side panel), and their fit to eq.~(\ref{eq:v_fit}).}
\label{fig:SpinProfile}
\end{figure*}

The analytic curves describing the spin profile for this model obey the following equation:
\begin{equation}
\langle \s_i (x_i) \rangle = v\left[ \left( 1-\frac{1}{2z^2} \right)\erf(z) + \frac{\exp\left(-z^2\right)}{\sqrt{\pi}z}\right]%
\label{eq:SpinProfile}
\end{equation}
where $ z = x_i \sqrt{\frac{2M}{R}}$.

Figure \ref{fig:SpinProfile} shows the excellent agreement between the numerical results (dots) obtained from simulations and the analytical curves (continuous lines) obeying eq.~(\ref{eq:SpinProfile}). All fits are done in the range $-25< x_i/a < 25$ in order to avoid systematic effects due to the boundaries of the system.

Following these results, we can state that in the broken symmetry phase we have a confirmation of the existence of topological excitations in the classical Heisenberg model as described in \cite{Delfino}.
In addition, one can extract the values of $v$ by fitting the numerical results to eq.~(\ref{eq:SpinProfile}). In the infinite-volume limit, this parameter is expected to scale as $v \propto (-t)^{\beta}$, where the critical exponent is expected to take the Gau{\ss}ian value $\beta = 1/2$, giving eq.~(\ref{eq:v_fit}). This is indeed what emerges from our numerical data as shown in the right-hand-side of figure \ref{fig:SpinProfile}.

\begin{equation}
  v = A_v \sqrt{1-\frac{T}{\Tc}}
  \label{eq:v_fit}
\end{equation}

Previous confirmation of the analytical predictions shown in \cite{Delfino} are present in \cite{XY} for the case of the XY model in three dimensions. In that work it is shown how the topological excitation enforced in the broken symmetry phase, also referred to as vortex, possesses a well-defined mass as the continuum limit is approached. This statement would imply a violation of Derrick's Theorem in scalar Quantum Field Theory.

In our numerical analysis, we investigated the possibility that the monopole-like topological excitation present in the classical Heisenberg model could have properties similar to the vortex found in the XY model by studying the scaling of the fitted values of $Ma$ as a function of the parameters of the theory: $L$, $R$, and $t$.
Following the interpretation of the topological field configuration as a particle excitation, $M$ should be interpreted as the mass of the particle, hence, independent from $R$ being the time extent of the lattice.
However, our fit results indicate that the parameter $Ma$ scales approximately proportionally to $R$. This observation questions the interpretation of the topological excitation as a physical particle.

\begin{figure*}[]
\centerline{\includegraphics[height=0.26\textheight]{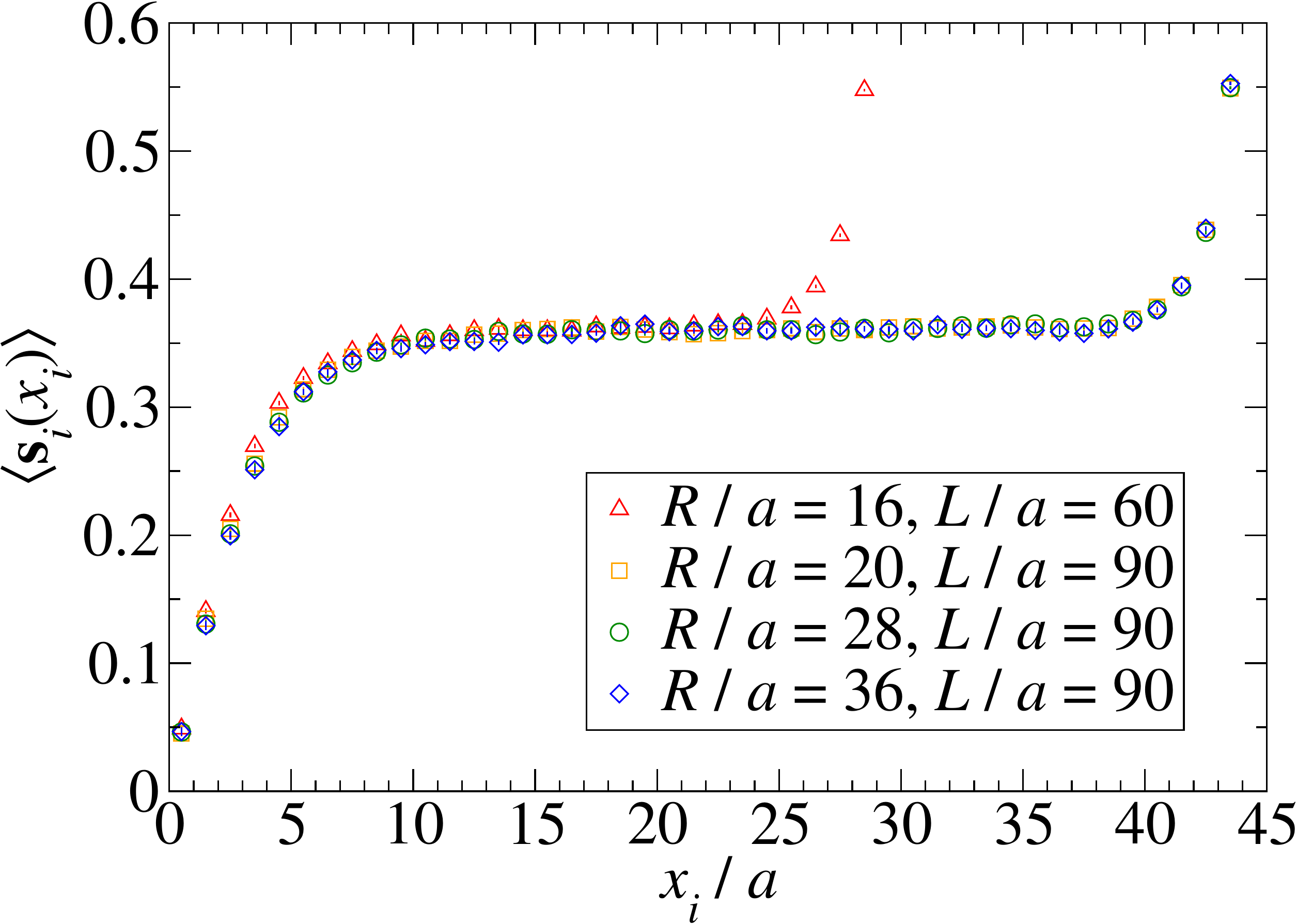}\hfill \includegraphics[height=0.26\textheight]{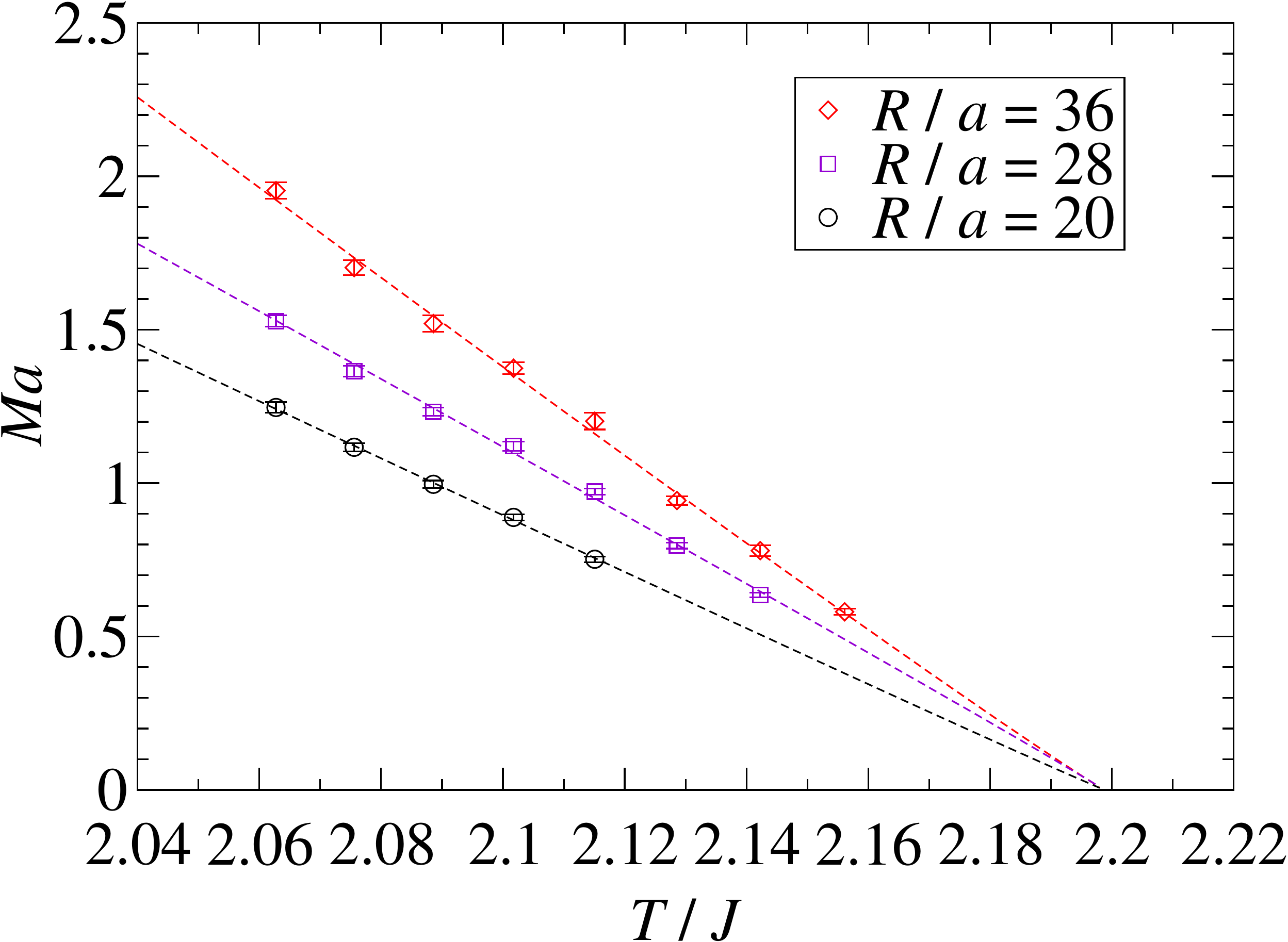}}
\caption{Left-hand-side panel: dependence of the spin profile $\langle \s_i (x_i) \rangle$ on the spatial coordinate $x_i/a$, as measured with respect to the center of the system, at a fixed temperature and for different values of the Euclidean-time extent of the lattice. All results shown in this plot were obtained from simulations with $T/J=2.062725$. Right-hand-side panel: temperature dependence of the $Ma$ parameter, extracted from the fits of our numerical data to eq.~(\ref{eq:SpinProfile}), for lattices with different values of $R/a$, and their fits to eq.~(\ref{eq:scalingFit}).} \label{fig:saturation}
\end{figure*}

Figure \ref{fig:saturation} shows how all data for the spin profile collapses onto the same curve, from which follows that $M \propto R$, or equivalently, that the topological excitation is characterised by an approximately constant $\mu = M/R$. If one think of $\mu$ as a physical quantity, then simulations at fixed $R$ but different values of $t$ should show $aM$ scaling like:
\begin{equation}
  a\mu R = aM \propto |t|^{2\nu}.
  \label{eq:scaling}
\end{equation}

By fitting our data to the function
\begin{equation}
  aM = A_M \left(1 - \frac{T}{\Tc}\right)^{E}
  \label{eq:scalingFit}
\end{equation}
we find results for $E$ very close to $1$, that is twice the value of $\nu$ predicted in the Gau{\ss}ian approximation of eq.~(\ref{eq:v_fit}).

\begin{table}[h]
\centering
\begin{tabular}{|c|l|l|l|}
\hline
$R/a$ & \multicolumn{1}{|c|}{$A_M$} & \multicolumn{1}{|c|}{$E$} & \multicolumn{1}{|c|}{$\redchisq$} \\
 \hline \hline
$20$ & $21.3(1.5)$ & $1.021(23)$ & $0.50$ \\
$28$ & $23.4(1.8)$ & $0.981(24)$ & $2.34$ \\
$36$ & $34.6(2.3)$ & $1.039(21)$ & $1.76$ \\
\hline
\end{tabular}
\caption{Results of the two-parameter fits of the data shown in the right-hand-side panel of figure~\ref{fig:saturation} to eq. (\ref{eq:scalingFit}).}                                                                                                               \label{tab:twonu_fit_results}
\end{table}

\section{Summary and Concluding Remarks}
From our study of the $O(3)$ spin model in four space-time dimensions, for the first time we achieve a numerical confirmation of Field Theoretical analytic prediction proposed by G. Delfino in \cite{Delfino}, as we observe the existence of a monopole-like excitation in the broken symmetry phase.
Moreover, we confirm the expected numerical results in the high-temperature phase.

Comparing our results with those in \cite{XY}, we investigated the possibility for the topological excitation considered in this model to violate Derrick's theorem.
From the results obtained in the broken symmetry phase, we conclude that the topological objects present in the classical Heisenberg model do not violate Derrick's theorem. An interesting consideration about the consequences of Derrick's theorem violation in the context of relativistic astrophysics is that it would imply the existence of bosonic stars consisting of particles that are excitations of real scalar fields \cite{BosonStar}.

Of course, there are alternatives that do violate the theorem, even though they are characterised by an intrinsic cutoff scale. Some examples of these systems do exist in condensed matter physics, and it is indeed in this setting that our findings might be particularly relevant. While our results are obtained in a classical statistical mechanics setting, one could interpret the model as a lattice regularisation of the corresponding quantum theory in three dimensions \cite{Book}. The possibility that this theory supports artificial monopole-like objects \cite{Monopoles} may have important applications in the efforts currently investigated towards quantum simulations \cite{QS}. For an experimental example of the work currently ongoing in this area see \cite{Manganese}.

Future studies on this model might include investigating the system in an out-of-equilibrium setting, as already explored in \cite{Panero}, which would allow us to study the non-equilibrium dynamics of the monopole excitation.

\bibliography{skeleton}


\end{document}